# Robustness of superconductivity to external pressure in high-entropy-alloy-type metal telluride AgInSnPbBiTe$_5$

Md. Riad Kasem[1], Yuki Nakahira[1], Hitoshi Yamaoka[2], Ryo Matsumoto[3,4], Aichi Yamashita[1], Hirofumi Ishii[5], Nozomu Hiraoka[5], Yoshihiko Takano[4], Yosuke Goto[1], Yoshikazu Mizuguchi[1]*

1. Department of Physics, Tokyo Metropolitan University, 1-1 Minami-Osawa, Hachioji 192-0397, Japan.

2. RIKEN SPring-8 Center, Sayo, Hyogo 679-5148, Japan

3. International Center for Young Scientists (ICYS), National Institute for Materials Science, Tsukuba, Ibaraki 305-0047, Japan

4. International Center for Materials Nanoarchitectonics (MANA), National Institute for Materials Science, Tsukuba, Ibaraki 305-0047, Japan

5. National Synchrotron Radiation Research Center, Hsinchu 30076, Taiwan

* Corresponding author: Y. Mizuguchi

**Email:** mizugu@tmu.ac.jp

## Abstract

High-entropy-alloy (HEA) superconductors are a new class of disordered superconductors. In this study, we investigate the robustness of superconducting states in HEA-type metal telluride ($M$Te; $M$ = Ag, In, Sn, Pb, Bi) under high pressure. PbTe exhibits a structural transition from a NaCl-type to an orthorhombic *Pnma* structure at low pressures, and further transitions to a CsCl-type structure at high pressures. When the superconductivity of the CsCl-type PbTe is observed, it is found that its superconducting transition temperature ($T_c$) decreases with pressure. However, in the HEA-type AgInSnPbBiTe$_5$, $T_c$ is almost independent of pressure, for pressures ranging from 13.0 to 35.1 GPa. This trend is quite similar to that observed in an HEA superconductor (TaNb)$_{0.67}$(HfZrTi)$_{0.33}$, which shows that the robustness of superconductivity to external pressure is a universal feature in HEA-type superconductors. To clarify the effects of the modification of the configurational entropy of mixing on the crystal structure, superconducting states, and electronic structure of $M$Te, electrical resistance measurements, synchrotron X-ray diffraction, and synchrotron X-ray absorption spectroscopy with partial fluorescence mode (PFY-XAS) for three $M$Te polycrystalline samples of PbTe, AgPbBiTe$_3$, and AgInSnPbBiTe$_5$ with different configurational entropies of mixing at the $M$ site were performed.

## Significance Statement

Motivated by the discovery of the robustness of superconductivity to extremely high pressure in a high-entropy-ally (HEA) superconductor (TaNb)$_{0.67}$(HfZrTi)$_{0.33}$, we studied the external pressure effects on the crystal structure, superconducting properties, and electronic structure of HEA-type metal telluride AgInSnPbBiTe$_5$. The robustness of superconductivity to external pressure was observed in the high-pressure phase (CsCl-type) of AgInSnPbBiTe$_5$, which is similar to that observed in (TaNb)$_{0.67}$(HfZrTi)$_{0.33}$. Our findings suggest that the robustness of superconductivity to external pressure is a universal feature in HEA-type superconducting materials. The combination of the effects of high pressure and HEA properties is expected to open a pathway to the development of new disordered superconductors with exotic properties.

## Introduction

The superconductivity of disordered materials has been extensively studied because of an observation made on the insulator-superconductor transition or disorder-induced superconductivity [1–3]. In addition, recent studies on $BiS_2$-based layered superconductors have shown the importance of controlling the local structural disorder to improve superconducting properties [4–7]. As a new category of disordered superconductors, high-entropy alloys (HEAs), which are alloys containing five or more elements with an atomic concentration between 5% and 35% [8,9], have been extensively studied, leading to the discovery of a wide range of HEA superconductors [10–12]. One of the notable features is the difference in characteristics of superconductivity among conventional metal or alloy superconductors, superconducting thin films, and HEA superconductors [11,13]. Furthermore, the robustness of superconductivity to extremely high pressure (HP), for pressures up to 190 GPa, in an HEA, $(TaNb)_{0.67}(HfZrTi)_{0.33}$, was observed [14]. The high configurational entropy of mixing ($\Delta S_{mix}$) reduces the Gibbs's free energy, and therefore stabilization of the phase is expected. The equation for calculating the Gibb's free energy is $\Delta G = \Delta H - T\Delta S_{mix}$, where $\Delta H$ is the enthalpy and $T$ is the absolute temperature. $\Delta S_{mix}$ can be calculated using the equation $\Delta S_{mix} = -R \sum_i c_i \ln c_i$, where, $R$ is the gas constant and $c_i$ is the atomic ratio of the element ($i$) [8]. The discovery of the robustness of superconductivity under an extremely high pressure was considered to be related to the high $\Delta S_{mix}$, but similar robustness of superconductivity under an extremely high pressure and a high transition temperature ($T_c$) of 19 K was observed in a low-entropy Nb-Ti alloy [15]. Therefore, the effects of $\Delta S_{mix}$ on the crystal structure, electronic structure, and superconducting properties of HEAs are still unclear. To address this issue, further investigation of HP effects on HEA-type materials is needed.

We recently developed superconducting HEA-type compounds, in which one of the crystallographic sites is alloyed with the criterion similar to HEAs [16]. Since compounds with two or more crystallographic sites have unique chemical bonds, various (novel) effects of the introduction of HEA site to electronic and structural properties would be expected, which is so-called cocktail effect in the field of HEAs. In the layered $BiS_2$-based superconductor $RE(O,F)BiS_2$, the rare-earth site ($RE$) was alloyed with five different $RE$ elements [17], and the improvement of superconducting properties (bulk nature of superconductivity) with increasing $\Delta S_{mix}$ was observed [18]. $T_c$ was not affected by $\Delta S_{mix}$, and therefore its insensitivity to $\Delta S_{mix}$ in the $BiS_2$-based superconductor is explained using a two-dimensional structure, since similar insensitivity was also observed in an $RE$123 cuprate with an HEA-type $RE$ site [19]. In contrast, $T_c$ of NaCl-type metal telluride, which is the target phase of this study, shows that its sensitivity to $\Delta S_{mix}$ of HEA-type systems is larger than that of metal tellurides with one or two constituent elements at the metal site [20,21].

On the basis of the facts described above, we considered that the metal telluride system is suitable for the discussion of the effects of high $\Delta S_{mix}$ on the crystal structure (phase stability) and superconducting properties under high pressure. Here, we studied the crystal structure, electronic structure, and superconducting properties of three tellurides ($M$Te; $M$ = Ag, In, Sn, Pb, Bi) with different $\Delta S_{mix}$ values at the $M$ site: PbTe ($\Delta S_{mix} = 0$), $AgPbBiTe_3$ ($\Delta S_{mix} = 1.10R$), and $AgInSnPbBiTe_5$ ($\Delta S_{mix} = 1.61R$). The pressure evolutions of the crystal structure and electronic transport properties were reported in previous studies [22–26]. Although PbTe is a semiconductor under ambient and low pressures, it undergoes metallization under high pressures, and a superconducting transition is observed for pressures above 15 GPa [22,23]. PbTe undergoes structural transitions as follows: cubic NaCl-type ($Fm$-3$m$) structure under low pressures, orthorhombic ($Pnma$) structure under moderate pressures, and cubic CsCl-type ($Pm$-3$m$) structure under HP [24–26]. There is a report on synthesis and thermoelectric properties of $AgPbBiTe_3$, but the structural and physical properties of $AgPbBiTe_3$ have not been reported [27]. In this study, we established a phase diagram of the crystal structures, as well as one for superconductivity versus pressure for both $AgPbBiTe_3$ and $AgSnInPbBiTe_5$. We also studied PbTe to examine the effects of configurational entropy of mixing. The pressure phase diagrams for the three tellurides are used to compare the crystal and electronic structures in detail. The results show that the pressure phase diagram of the crystal structure is largely modified by the effect of $\Delta S_{mix}$. Furthermore, we found that the pressure dependence of $T_c$ in the HP phase (CsCl-type phase) exhibits a clear difference among the three compounds. In the HP phase, the $T_c$ for PbTe decreases with pressure, but that for HEA-type AgInSnPbBiTe$_5$ exhibits a flat dependence, which suggests that the superconducting state of AgInSnPbBiTe$_5$ is robust to pressure. Our present findings show an analogical conclusion in the case of (TaNb)$_{0.67}$(HfZrTi)$_{0.33}$ [14] and suggest that the HEA effects universally enhance the robustness of superconductivity under high pressure.

## Results

### Electrical resistance

Figure 1A shows the temperature dependence of the electrical resistance of AgInSnPbBiTe$_5$ measured using the conventional four-probe method at ambient pressure. $T_c^{zero}$ was 1.8 K, which is slightly lower than that reported in previous research [20]. We measured the electrical resistance (at National Institute for Materials Science) after several days since the sample was synthesized by HP annealing (at Tokyo Metropolitan University), and a slight aging effect of $T_c$ was noticed in $M$Te superconductors synthesized under high pressure [28,29]. We consider that the slight decrease in $T_c$ can be understood by the change in the internal strains because the XRD patterns do not show a remarkable change after aging in the HP-synthesized $M$Te samples. Therefore, we continued to perform resistance measurements under high pressure using a diamond-anvil cell (DAC).

Figure 1B shows the temperature dependences of the electrical resistance of AgInSnPbBiTe$_5$ measured under various pressures with a DAC. Normal-state resistance decreases with pressure, for pressures up to 9.3 GPa and increases with pressure for pressures above 15.5 GPa. An onset of the superconducting transition was observed at 2 K under 6.5 GPa, and the zero-resistance state was observed at $T_c^{zero}$ = 2 K under 10.2 GPa. As shown in Fig. 2A, the $T_c^{onset}$ monotonously increased with pressure until it reached 13.0 GPa, and then it became insensitive to pressure for pressures ranging from 13.0 GPa to 35.1 GPa ( P = 35.1 GPa is the chosen maximum pressure for the experiment). The results indicate that the $T_c$ in the CsCl-type structure of AgInSnPbBiTe$_5$ is independent of applied pressure, while lattice constant decreases with pressure. The highest $T_c^{zero}$ and $T_c^{onset}$ observed in the experiment on AgInSnPbBiTe$_5$ are 4.5 K and 5.3 K, respectively. See Fig. S1 for the determination criterion for $T_c^{onset}$.

To compare the pressure evolution of $T_c$ in AgInSnPbBiTe$_5$ with that of PbTe and AgPbBiTe$_3$, the pressure dependences of $T_c^{onset}$ for PbTe and AgPbBiTe$_3$ were examined (Figs. S2, S3), and the resulting $T_c$-$P$ plots are shown in Fig. 2B. PbTe is a semiconductor at ambient and low pressures, but exhibits a pressure-induced superconducting transition above 17 GPa in the CsCl-type structure [22]. At higher pressures, $T_c$ of PbTe monotonously decreases with pressure. In the case of AgPbBiTe$_3$, superconductivity was observed at $P$ > 2.6 GPa, and at this instance $T_c^{onset}$ reached 6.5 K. The $T_c$ for AgPbBiTe$_3$ slightly decreases at high pressures. In AgInSnPbBiTe$_5$, superconductivity is observed at low pressures as well because the low-pressure phase, having a NaCl-type structure, itself is a metal and shows superconductivity under ambient pressure (Figs. 1A and 2A). As demonstrated in the next section, the crystal-structure type under high pressure is CsCl-type for all the compounds. However, the trend of the pressure dependences of $T_c$ in the CsCl-type structure exhibit a clear difference among PbTe, AgPbBiTe$_3$, and AgInSnPbBiTe$_5$. The main findings of this study are that the robustness of superconductivity to pressure in HEA-type AgInSnPbBiTe$_5$ is similar to that observed in (TaNb)$_{0.67}$(HfZrTi)$_{0.33}$ [14]. To validate the conclusion, we investigated the pressure evolutions of the crystal structure and the electronic structure for those $M$Te samples under high pressure.

### Crystal structure

Figures 3A–3C show the pressure-dependent synchrotron X-ray diffraction (SXRD) patterns for the PbTe, AgPbBiTe$_3$, and AgInSnPbBiTe$_5$ samples, respectively. For all the SXRD patterns, we performed the Rietveld refinement to confirm the structural type and to evaluate the lattice constant. See Tables S1–S3 and Figs. S4–S6 for details on refinements. On the basis of the refinement results, we established structural phase diagrams under high pressure (Figs. 3D–3F) by plotting the pressure dependence of volume per unit formula (*Z*). For PbTe, the structural transition from NaCl-type to *Pnma* occurs at around 6.80 GPa, and the second transition to CsCl-type takes place at 14.28 GPa. The transition gradually occurred, hence the phase diagram contains mixed phases. The results on PbTe are consistent with the previous work by Li *et al.* [24]. For AgPbBiTe$_3$ and AgInSnPbBiTe$_5$, similar phase diagrams were obtained, where the NaCl-type structure is stabilized up to ~10 GPa, and the *Pnma* phase is suppressed. The pressure where the CsCl-type phase is induced is common to the case of PbTe. In all the structural types including the CsCl-type phase, the lattice volume continuously decreases with pressure. Although the difference in the stability of the NaCl-type and *Pnma* structures may be related to the difference in lattice volume at ambient pressure, we consider that the *Pnma* phase is suppressed, and the NaCl-type phase is stabilized by the effect of alloying at the *M* site. We note that configurational entropy of mixing does not affect the structure of the CsCl-type phase, and lattice volume of the CsCl-type phase commonly decreases with pressure in three *M*Te sample.

### Electronic structure

To examine the effects of pressure and HEA states on the electronic structure, we performed X-ray absorption spectroscopy with partial fluorescence mode (PFY-XAS) for PbTe and AgInSnPbBiTe$_5$. See Fig. S7 for pressure dependences of spectra, and analysis results on the Pb-$L_3$ and Bi-$L_3$ spectra. In general, the absorption spectra at the Pb-$L_3$ absorption edge are similar to those at the Bi-$L_3$ absorption edge. On the basis of analogically referring to other Pb- or Bi-containing compounds [30,31], we analyzed the spectra by assuming several peaks. An example of the fit for the PFY-XAS spectra at 27 GPa is shown in Fig. 4A. The PFY-XAS spectra were fitted by assuming some Voigt functions with an arctan-like background [31]. In this study, we focus on the peaks of P1, P2, and P3, where peak P1 could be assigned as a dipole transition of the $2p_{3/2}$ electron into the $6s$ state, and the peaks shown as P2 and P3 correspond to $6d$ states of $t_{2g}$ and $e_g$, respectively [32,33]; we measure $2p_{3/2} \rightarrow nd$ ($n > 6$) transitions at the Pb $L_3$ absorption edge. There is a $p$ density of states ($p$ DOS) above the Fermi level, however, we mainly observe the dipole-allowed transitions of Pb $2p$–$6s$ and Pb $2p$–$6d$, and therefore, the observed spectra do not reflect the Pb $6p$ DOS spectroscopically. The absorption spectra reflect the empty DOS above the Fermi level generally, with a core hole in the final state.

For PbTe with a NaCl-type structure, there is a narrow band gap, and the $p$ orbitals of Pb and Te near the Fermi energy are hybridized. See Fig. S8 for the calculated DOS for PbTe [34]. The valence band is mainly composed of Te $5p$ orbitals, and also contains the contribution from Pb $6p$ and $6s$, while the conduction band is mainly composed of Pb $6p$ orbitals, but also contains Te $5p$ contributions. For PbTe with a CsCl-type structure, band gap is totally closed, and the DOS near the Fermi energy is explained by the contributions from Pb $6p$ and $6s$ orbitals, as well as the Te $5p$ orbitals. Therefore, the pressure evolution of the 6s states which could be resolved in our high-resolution spectroscopy, may play an important role on the closing of the band gap as well as the emergence of superconductivity.

The pressure dependences of the intensity and the energy of peak P1 in PbTe is shown in Fig. 4B. The intensity of peak P1 in Fig. 4B gradually increases with pressure, up to about 5 GPa. The increase in the intensity of P1 indicates an increase in the amount of holes in the Pb 6$s$ states, which indicates a modification of the band structure. In the middle-pressure phase, (between 5-15 GPa) the intensity of P1 does not show a significant change, while it increases remarkably in the HP phase (P > 15 GPa). We note that PbTe with a NaCl-type or *Pnma* structure is a semiconductor with a band gap at the low-pressure regime, and we also note that the metallic phase is induced in a CsCl-type structure for pressures above 15 GPa [22,34,35]. The increase of the intensity of P1 corresponds with the increase of the unoccupied 6$s$ states of the Pb. This may correlate with the emergence of the superconductivity after the closing of the band gap at $P > 15$ GPa.

On the other hand, the energy of peak P1 shifts to a lower incident energy until it reaches a pressure level of 5 GPa. The incident energy and the intensity do not change in the middle-pressure range of 5–17 GPa, and they start decreasing again for pressures above 18 GPa as shown in Fig. 4B. The shift of the energy of peak P1 to a lower incident energy is explained by the upward shift of the Fermi level or change in the DOS at the Fermi level. Theory suggests that the energy shift of peak P1 may be influenced by the reduction of the band gap [35] and the theoretical band gap is in the same order as the energy shift of P1 at 5 GPa.

The intensity of P3 (Pb 6$d$ DOS, $e_g$) shows a trend which is similar to the intensity of P1. The intensity of P2 (Pb 6$d$ DOS, $t_{2g}$) on the other hand decreases with increasing pressure at $P > 17$ GPa (Fig. 4C). It is interesting that there is a large change in the electronic structure for pressures above 20 GPa, but the crystal structure still retains its CsCl-type structure in this pressure range. In PbTe, superconductivity suddenly appears above 18 GPa, and $T_c$ decreases with pressure monotonically [22]. The present result possibly suggests that the change in the electronic structure is not favorable for the superconductivity of *M*Te, when it has transitioned to the CsCl-type structure.

We also measured the PFY-XAS spectra at the Pb-$L_3$ absorption edges for AgInSnPbBiTe$_5$ as shown in Fig. S7. We observed similar trends in the pressure dependence of the electronic structures as those observed for PbTe. An example of the fit at 28.8 GPa is shown in Fig. 4D. The analysis results on P1, P2, and P3 are plotted in Figs. 4E and 4F. Figure 4E shows a gradual increase of the P1 intensity with pressure, which is similar to the case of PbTe. The trend of P2 is also similar for the entire pressure range, and that of P3 is basically similar between PbTe and AgInSnPbBiTe$_5$. The PFY-XAS spectra at the Bi-$L_3$ absorption edge were also taken, and the analysis results are summarized in Fig. S7. In AgInSnPbBiTe$_5$ the pressure-induced change in the electronic structure seems to be common for Bi and Pb sites; the detailed results are shown under the Supporting Information section. In conclusion, the electronic structures of PbTe and AgInSnPbBiTe$_5$ show a similar pressure dependence, even though the structure of PbTe does not change much in the middle-pressure range (*Pnma* + CsCl phase), which disappears in AgInSnPbBiTe$_5$. Therefore, the difference in the robustness of superconductivity to pressure in the CsCl-type phase between PbTe and AgInSnPbBiTe$_5$ cannot be explained by the pressure evolutions of crystal and electronic structures.

## Discussion

From the structural viewpoint, the impact of the introduction of an HEA site is the suppression of the middle-pressure phase with a *Pnma* structure. In other words, the low-pressure phase with a NaCl-type structure is stabilized up to a higher pressure of $P > 10$ GPa in AgInSnPbBiTe$_5$, whereas the NaCl-type phase disappears at ~5 GPa for PbTe. Interestingly, the trend of lattice constant in the CsCl-type structure under HP is quite similar for PbTe and AgInSnPbBiTe$_5$. However, as revealed in Fig. 2B, the pressure dependences of $T_c$ clearly differ since $T_c$ decreases with pressure for PbTe but does not change largely in the CsCl-type phase for AgInSnPbBiTe$_5$. Furthermore, from the electronic-structure viewpoint, we cannot find a clear correlation between the robustness of superconductivity to pressure and the changes in electronic structure under high pressure. Although there is a possibility of the difference of the contribution of the 6$p$ states, which could not be measured in our spectra, to the conduction band to increase the number of the carriers in AgInSnPbBiTe$_5$.

The results show that the pressure phase diagram of the crystal structure is largely modified by the effect of $\Delta S_{mix}$. In contrast, the electronic structures are not sensitive to the effect of $\Delta S_{mix}$. To understand the HEA effects on structural and electronic properties for $M$Te under high pressure, further studies using various probes are needed. However, commonality on the robustness of superconductivity to external pressure in the superconducting HEA (TaNb)$_{0.67}$(HfZrTi)$_{0.33}$ [14] and the HEA-type metal telluride AgInSnPbBiTe$_5$ would be demonstrating the universal characteristics of superconductivity in HEA-type materials. Thus, the present results propose that the combination of HP and HEA effects will open a new pathway to the development of new disordered superconductors with exotic superconducting states.

## Materials and Methods

The polycrystalline sample of PbTe was synthesized by the solid-state reaction of Pb (99.9%) and Te (99.999%) at 900 ºC. To obtain a pellet for resistance measurements, pelletizing and second annealing were performed. The polycrystalline samples of AgPbBiTe$_3$ and AgInSnPbBiTe$_5$ were synthesized using an HP synthesis method where the pressure was kept below 3 GPa, and the temperature kept at 500 ºC for 30 minutes as described in Ref. 20. The precursor powders of AgPbBiTe$_3$ and AgInSnPbBiTe$_5$ were synthesized by a solid-state reaction of Ag powders (99.9%) and grains of In (99.99%), Sn (99.999%), Pb (99.9%), Bi (99.999%), and Te (99.999%) at 800 ºC, with the nominal compositions.

The electrical resistance measurements were performed at ambient pressure using the conventional four-probe method on a GM refrigerator system. Resistance measurements under high pressure were performed on polycrystalline powder on a Physical Property Measurement System (Quantum Design) using an originally designed diamond anvil cell (DAC) with boron-doped diamond electrodes [36–38]. The sample was placed on the boron-doped diamond electrodes in the center of the bottom anvil. The surface of the bottom anvil, except for the sample space and electrical terminal, were covered with the undoped diamond insulating layer. The cubic boron nitride powders with ruby manometer were used as a pressure-transmitting medium. The applied pressure was estimated by the fluorescence from ruby powders [39] and the Raman spectrum from the culet of top diamond anvil [40] using an inVia Raman microscope (RENISHAW).

Pressure dependences of the synchrotron X-ray powder diffraction (SXRD) patterns were measured at BL12B2, SPring-8, using a 3-pin plate diamond anvil cell (DAC, Almax easyLab Industries) with a CCD detection system at room temperature (~ 293 K). Culet size of the diamond anvil was 0.4 mm with a stainless-steel gasket. We took an arrangement of both incoming and outgoing x-ray beams passing through the diamonds with incident photon energy of 18 keV. A two-dimensional image of the CCD system was integrated using the FIT2D program [41]. Silicone oil was used as a pressure-transmitting medium, and pressure was monitored using the ruby fluorescence method. [42] The SXRD data was analyzed through Jana 2006 software [43] using the Rietveld method.

The pressure dependence of the high-resolution X-ray absorption spectra was measured at beamline BL12XU, SPring-8. Membrane-controlled DACs with a 0.3 mm culet and that with a 0.4 mm culet were used for PbTe and AgInSnPbBiTe$_5$, respectively, and silicone oil was used as a pressure-transmitting medium. Beryllium gaskets with a 3 mm diameter were pre-indented at the center. The thickness was approximately 67 μm for PbTe and 31 μm for AgInSnPbBiTe$_5$, and the diameters of the sample chamber in the gaskets were approximately 110 μm for PbTe and 140 μm for AgInSnPbBiTe$_5$. We employed X-ray absorption spectroscopy (XAS) with a partial fluorescence mode (PFY-XAS), which has an advantage of a higher resolution as compared to that of normal XAS [44,45]. We used the Be gasket in plane geometry where both incoming and outgoing x-ray beams passed through the Be gasket. A Johann-type spectrometer equipped with a spherically bent Si(555) analyzer crystal (radius of ~1 m), and a Si solid state detector were used to analyze the Bi $L\alpha_1$ (10.839 keV, $3d_{5/2}$-$2p_{3/2}$) emission at the Bi $L_3$ absorption edge, and Pb $L\alpha_1$ (10.551 eV, $3d_{5/2}$-$2p_{3/2}$) emission at the Pb $L_3$ absorption edge [46]. The incident beam is focused at 17 μm x 40 μm by the K-B mirror located at the sample position.

**Acknowledgments**

The measurements of the XRD patterns and the PFY-XAS spectra under pressure were performed BL12XU, and BL12B2, SPring-8 under SPring-8 Proposal No. 2020A4269 (corresponding to Proposal No. 2021-1-09-1 of NSRRC). The authors thank O. Miura for his supports in experiments. This work was partially supported by Grant-in-Aid for Scientific Research (KAKENHI) (Nos. 18KK0076, 21K18834, 21H00151) and Tokyo Metropolitan Government Advanced Research (H31-1). We thank Editage for English correction of the manuscript.

**Figures**

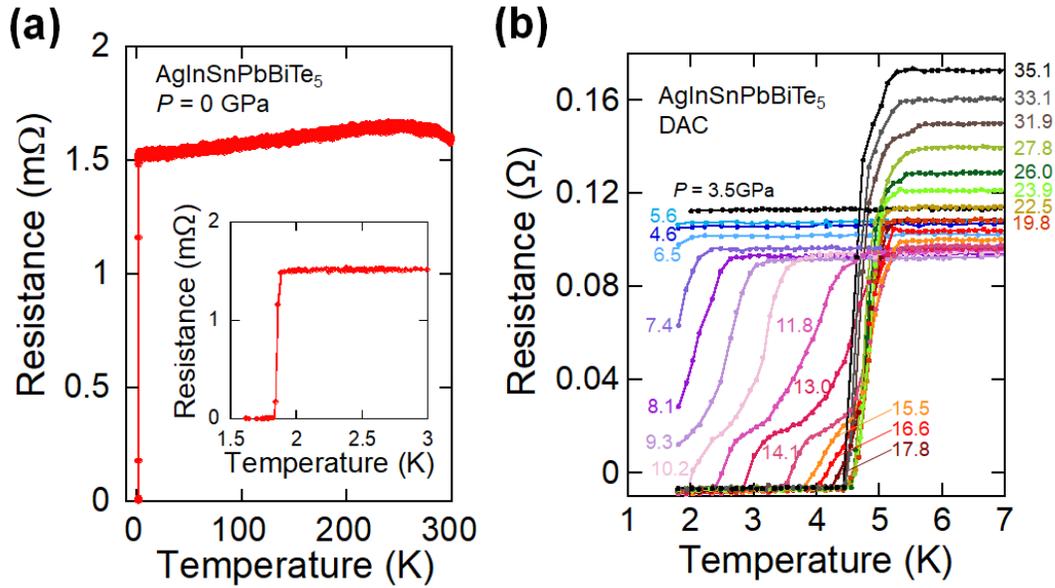

Figure 1. (a) Temperature dependence of electrical resistance of AgInSnPbBiTe$_5$ measured using the conventional four-probe method. (b) Temperature dependences of electrical resistance of AgInSnPbBiTe$_5$ measured using a diamond anvil cell (DAC).

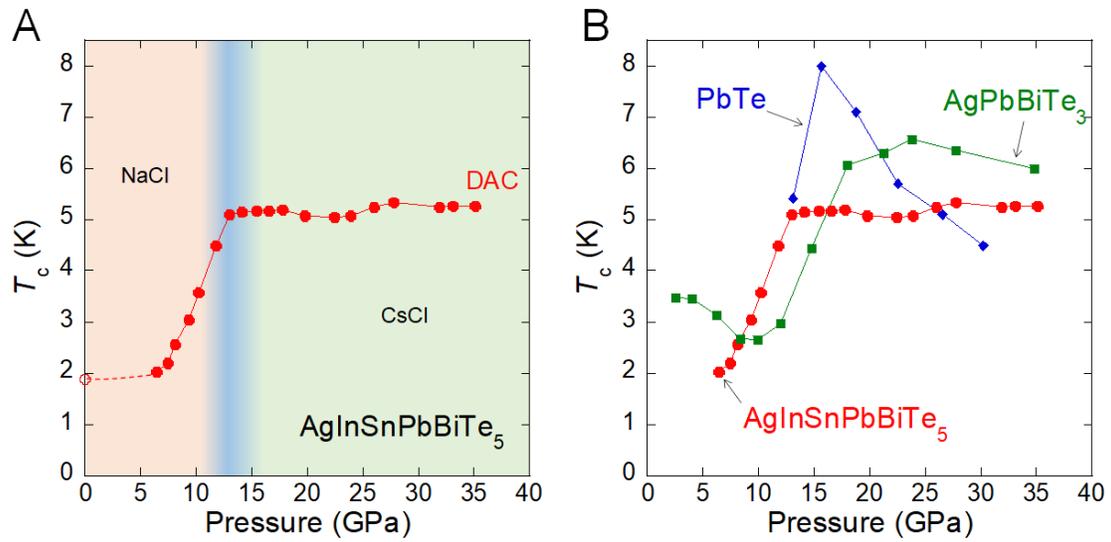

Figure 2. (A) Pressure dependence of $T_c$ of AgInSnPbBiTe$_5$. Open and filled circles indicate the data taken without a pressure cell, and the data that was measured with DAC. The structural types are indicated according to structural analyses in Fig. 3D. (B) Pressure dependences of Tc for Pb, AgPbBiTe$_3$, and AgInSnPbBiTe$_5$.

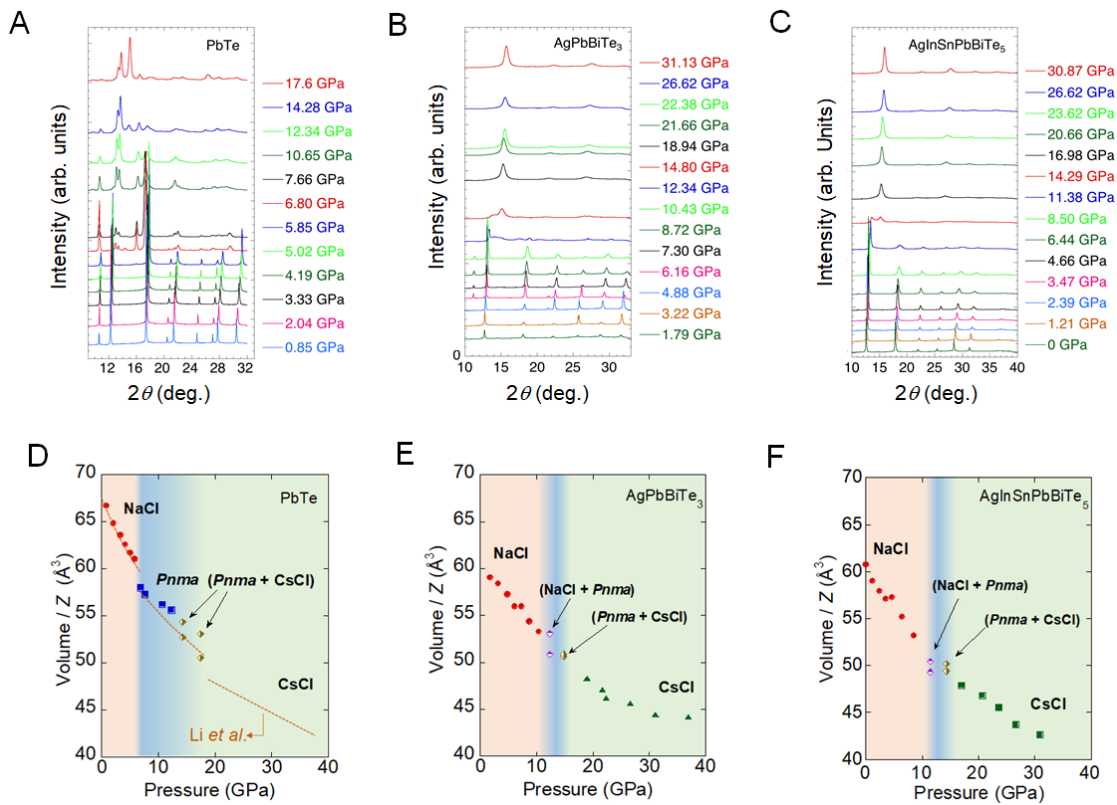

Figure 3. (A–C) SXRD patterns for PbTe, AgPbBiTe$_3$, and AgInSnPbBiTe$_5$. Note that the baseline height of the XRD pattern at each pressure scales to the pressure. (D–F) Lattice volumes divided by $Z$ (chemical formula sum in a unit cell) for PbTe, AgPbBiTe$_3$, and AgInSnPbBiTe$_5$ are plotted as a function of pressure. In Fig. 3D, the analysis results reported in Ref. 24 (Li *et al.*) are represented by orange lines. Structural types are shown in the figures.

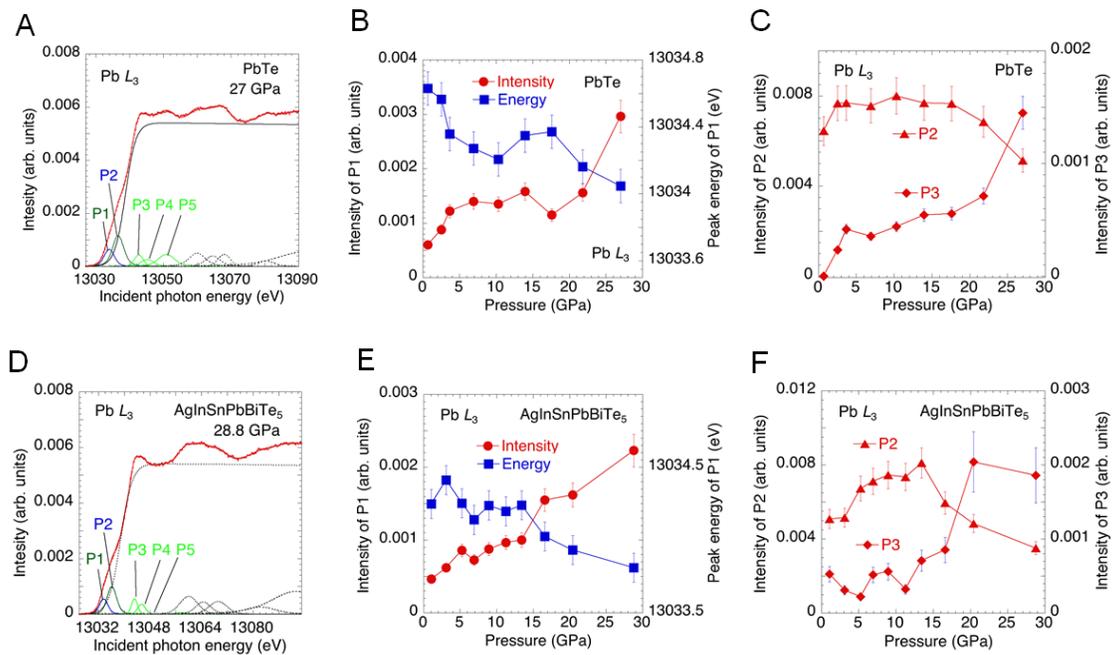

Figure 4. (A) Typical Pb-$L_3$ PFY-XAS spectrum and an example of the fit to the spectrum for PbTe at 27 GPa. (B) Pressure dependence of the intensity and the energy of the peak P1 for PbTe. (C) Pressure dependence of the intensity of the peaks P2 and P3 for PbTe. (D) Typical Pb-$L_3$ PFY-XAS spectrum and an example of the fit to the spectrum for AgInSnPbBiTe$_5$ at 28.8 GPa. (E) Pressure dependence of the intensity and the energy of the peak P1 for AgInSnPbBiTe$_5$. (F) Pressure dependence of the intensity of the peaks P2 and P3 for AgInSnPbBiTe$_5$.

*Supporting Information*

Table S1. Summary of the major and minor phases and the lattice constants revealed by the refinements for PbTe.

| | | | | PbTe | | | | |
|---|---|---|---|---|---|---|---|---|
| P (GPa) | Major phase | Minor phase | Major phase | | | Minor phase | | |
| | | | a (Å) | b (Å) | c (Å) | a (Å) | b (Å) | c (Å) |
| 0.85 | NaCl | - | 6.41241(7) | | | | | |
| 2.04 | NaCl | - | 6.37689(3) | | | | | |
| 3.33 | NaCl | - | 6.33345(3) | | | | | |
| 4.19 | NaCl | - | 6.30221(6) | | | | | |
| 5.02 | NaCl | - | 6.27128(4) | | | | | |
| 5.85 | NaCl | *Pnma* | 6.24916(5) | | | 8.149(2) | 4.5447(11) | 6.2867(8) |
| 6.80 | *Pnma* | - | 8.113(3) | 4.5391(2) | 6.2977(10) | | | |
| 7.66 | *Pnma* | - | 8.009(2) | 4.5675(2) | 6.263(2) | | | |
| 10.65 | *Pnma* | - | 7.931(2) | 4.5432(4) | 6.2379(12) | | | |
| 12.34 | *Pnma* | - | 7.9041(11) | 4.5269(5) | 6.2161(7) | | | |
| 14.28 | *Pnma* | CsCl | 7.849(9) | 4.4896(6) | 6.1670(4) | 3.7498(8) | | |
| 17.60 | CsCl | *Pnma* | 3.69712(10) | | | 7.8775(13) | 4.4080(7) | 6.1121(3) |

Table S2. Summary of the major and minor phases and the lattice constants revealed by the refinements for AgPbBiTe$_3$.

| | | | | AgPbBiTe$_3$ | | | | |
|---|---|---|---|---|---|---|---|---|
| P (GPa) | Major phase | Minor phase | Major phase | | | Minor phase | | |
| | | | a (Å) | b (Å) | c (Å) | a (Å) | b (Å) | c (Å) |
| 1.79 | NaCl | - | 6.18099(10) | | | | | |
| 3.22 | NaCl | - | 6.15772(9) | | | | | |
| 4.88 | NaCl | - | 6.11365(10) | | | | | |
| 6.16 | NaCl | - | 6.07139 (10) | | | | | |
| 7.30 | NaCl | - | 6.07131(11) | | | | | |
| 8.72 | NaCl | - | 6.01276(9) | | | | | |
| 10.43 | NaCl | - | 5.97220(13) | | | | | |
| 12.34 | NaCl | *Pnma* | 5.8814(2) | | | 8.985(2) | 3.6292(2) | 6.493(2) |
| 14.80 | *Pnma* | CsCl | 7.5294(9) | 4.4553(6) | 6.0653(9) | 3.6995(4) | | |
| 18.94 | CsCl | - | 3.6377(2) | | | | | |
| 21.66 | CsCl | - | 3.6119(2) | | | | | |
| 22.38 | CsCl | - | 3.58387 (13) | | | | | |
| 26.62 | CsCl | - | 3.5737(2) | | | | | |
| 31.13 | CsCl | - | 3.53490(10) | | | | | |
| 36.92 | CsCl | - | 3.53489(11) | | | | | |

Table S3. Summary of the major and minor phases and the lattice constants revealed by the refinements for AgInSnPbBiTe$_5$.

| P (GPa) | Major phase | Minor phase | Major phase a (Å) | Major phase b (Å) | Major phase c (Å) | Minor phase a (Å) | Minor phase b (Å) | Minor phase c (Å) |
|---|---|---|---|---|---|---|---|---|
| 0.00 | NaCl | - | 6.2386(2) | | | | | |
| 1.21 | NaCl | - | 6.18054(7) | | | | | |
| 2.39 | NaCl | - | 6.14222(7) | | | | | |
| 3.47 | NaCl | - | 6.11298(13) | | | | | |
| 4.66 | NaCl | - | 6.0782(2) | | | | | |
| 6.44 | NaCl | - | 6.0433(2) | | | | | |
| 8.50 | NaCl | - | 5.9682(3) | | | | | |
| 11.38 | NaCl | Pnma | 5.86438(7) | | | 7.918(2) | 4.2683(5) | 5.836(2) |
| 14.29 | Pnma | CsCl | 9.2478 (11) | 3.6954(3) | 5.8745 (4) | 3.6694 (2) | | |
| 16.98 | CsCl | - | 3.6309 (2) | | | | | |
| 20.66 | CsCl | - | 3.60484(13) | | | | | |
| 23.62 | CsCl | - | 3.57124(9) | | | | | |
| 26.62 | CsCl | - | 3.52154(11) | | | | | |
| 30.87 | CsCl | - | 3.49282(8) | | | | | |

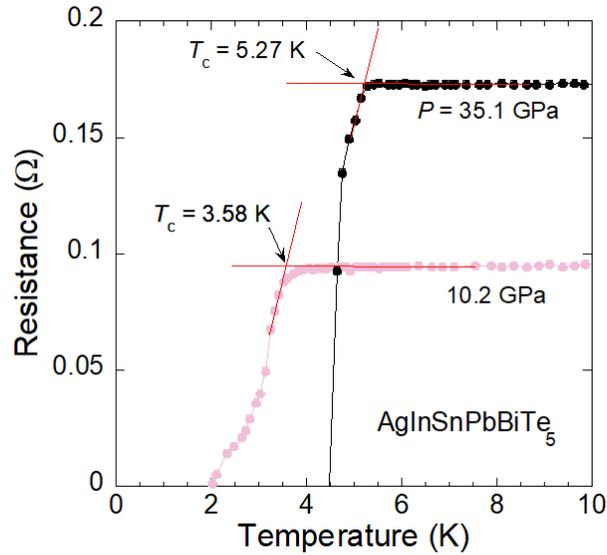

Fig. S1. Estimation of $T_c$ from the DAC resistance data of AgInSnPbBiTe$_5$. $T_c$ was estimated form a cross point of two (red) lines.

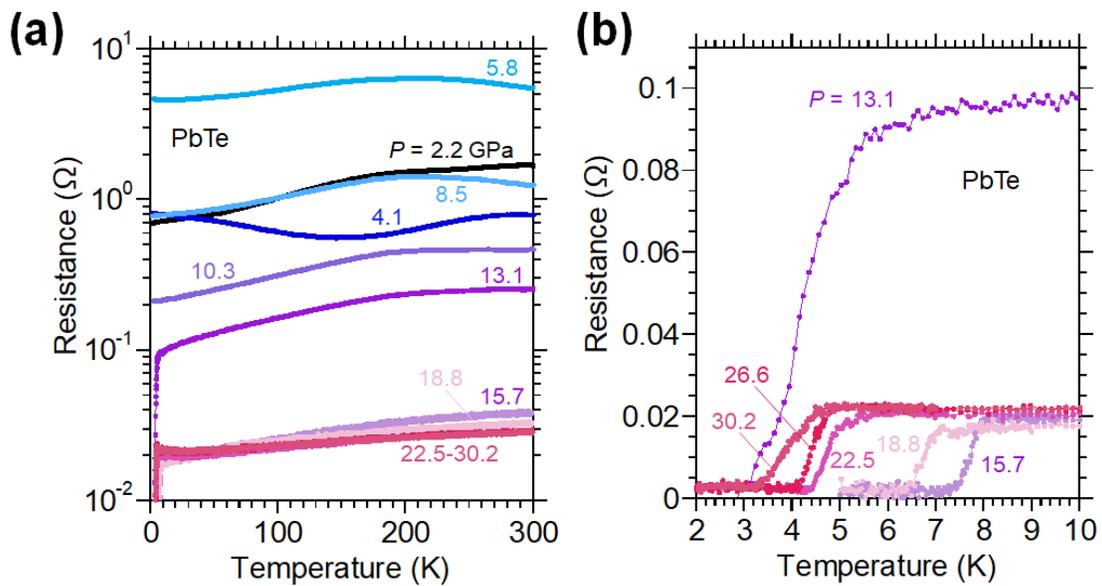

Fig. S2. Temperature dependences of electrical resistance for PbTe under high pressures.

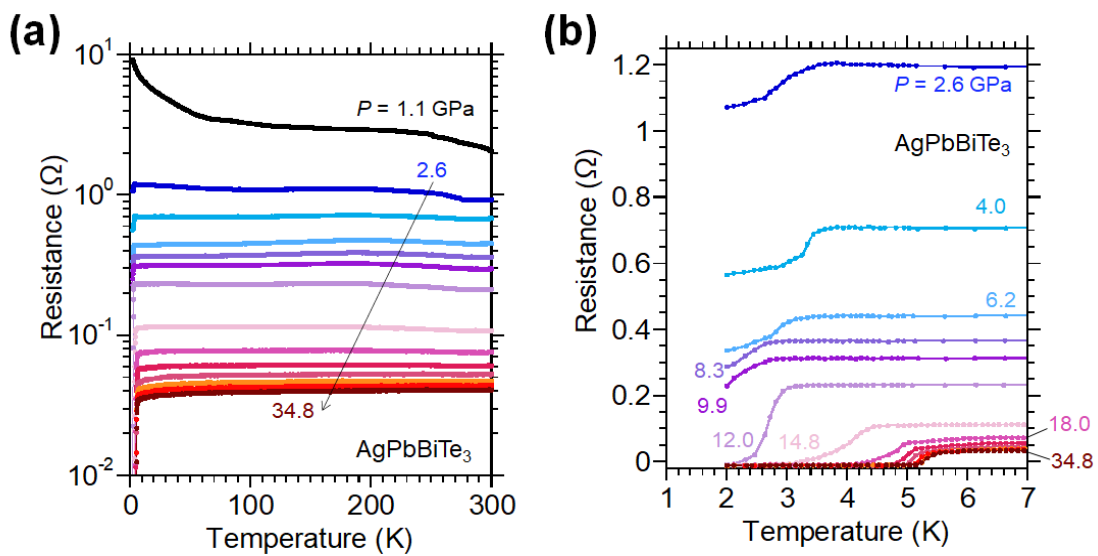

Fig. S3. Temperature dependences of electrical resistance for AgPbBiTe$_3$ under high pressures.

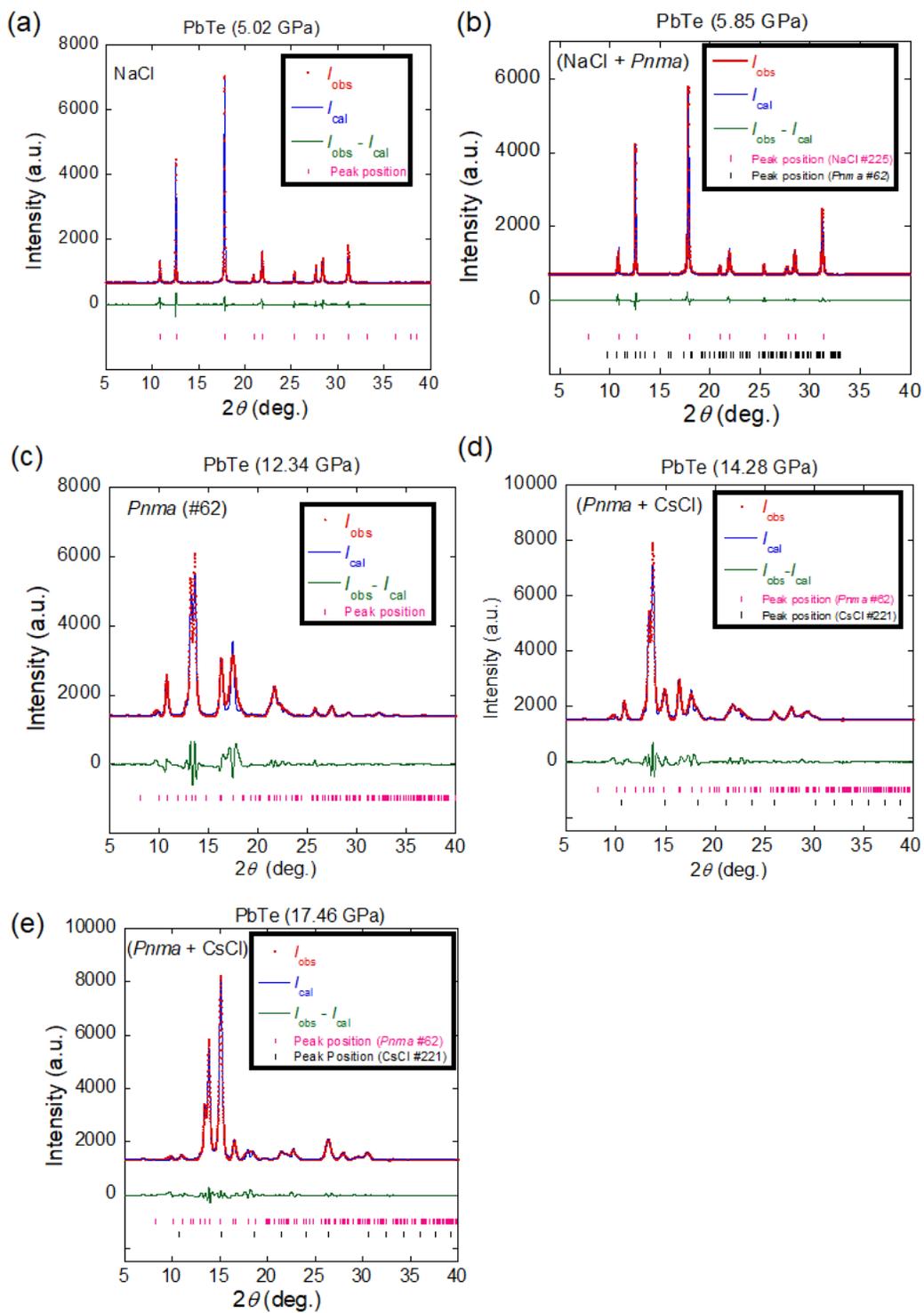

Fig. S4. Typical refinement results for PbTe.

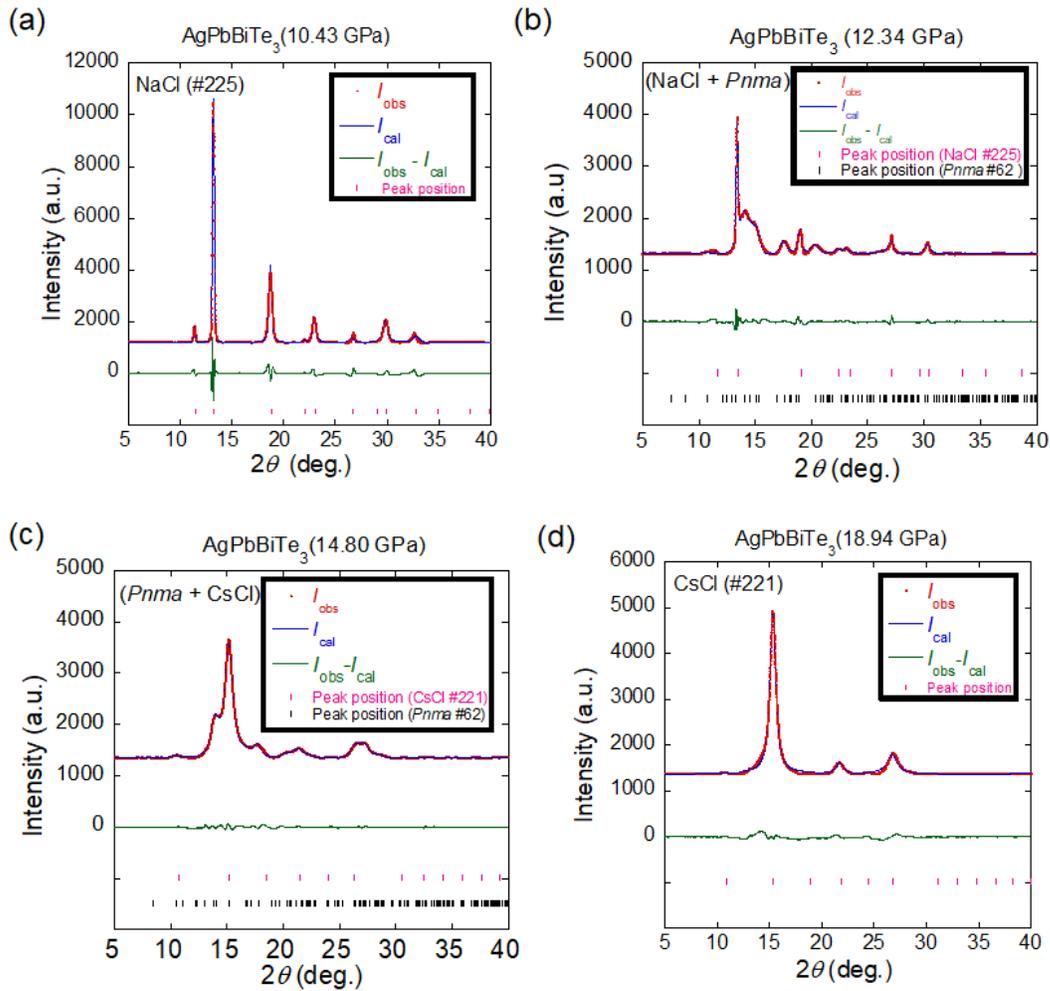

Fig. S5. Typical refinement results for AgPbBiTe$_3$.

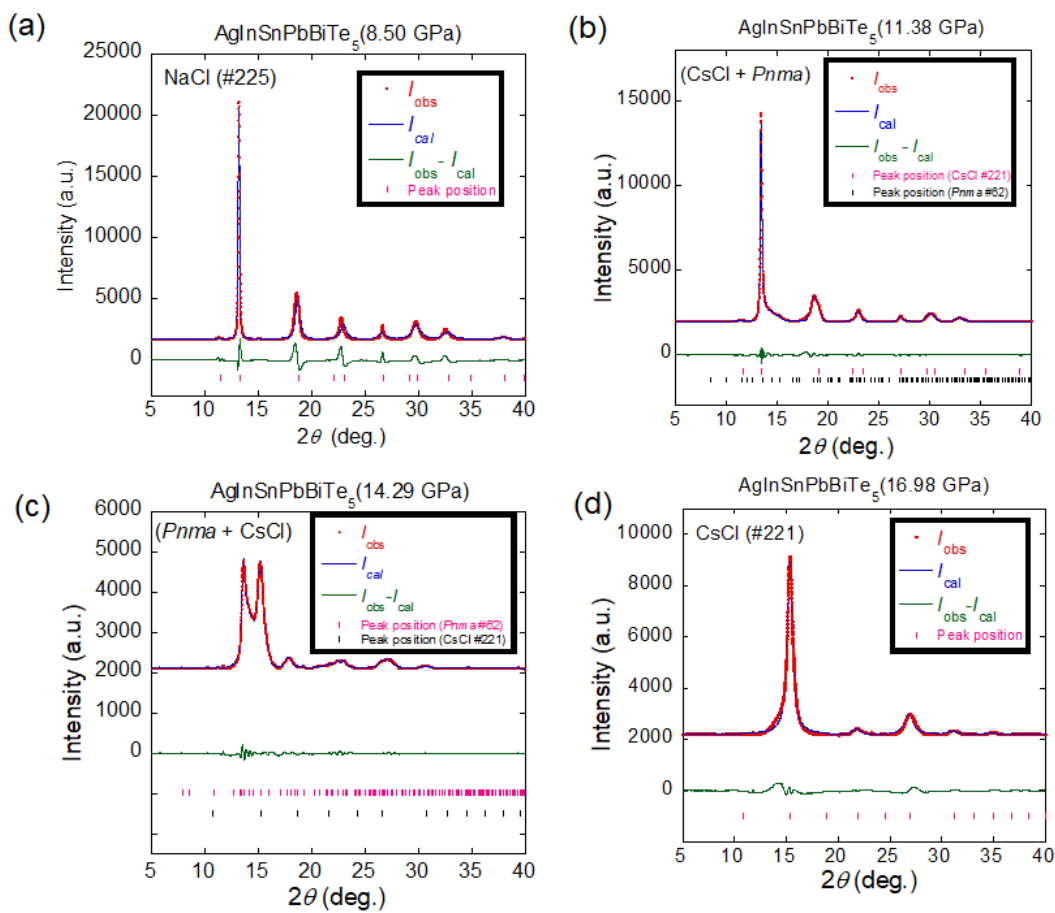

Fig. S6. Typical refinement results for AgPbBiTe$_3$.

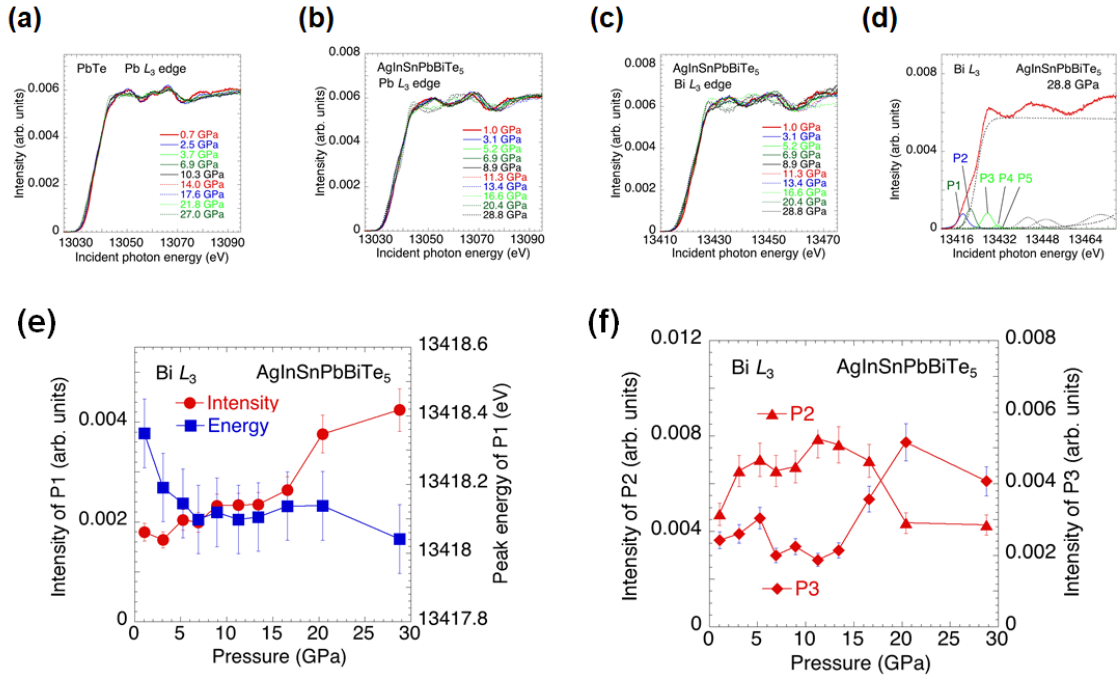

Fig. S7. (a,b) Pressure dependence of the PFY-XAS spectra at the Pb-$L_3$ absorption edge for PbTe, and AgInSnPbBiTe$_5$. (c) Pressure dependence of the PFY-XAS spectra at the Bi-$L_3$ absorption edge for AgInSnPbBiTe$_5$. (d) An example of the Bi-$L_3$ PFY-XAS spectrum at 28.8 GPa. (e,f) Pressure dependence of the intensity and the energy for the peaks P1, P2, and P3 obtained from the Bi-$L_3$ PFY-XAS data for AgInSnPbBiTe$_5$.

As shown in Fig. S7, the intensity of P1 increases with pressure in the whole pressure range measured in this study. Both intensities of P2 and P3 show a trend to increase in the NaCl-type structure phase up to ~15 GPa. The intensity of P2 decreases with pressure with pressure above ~15 GPa in the in the CsCl-type structure phase, while that of P3 decreases. The energy of P1 show a trend to decrease with pressure in both Bi and Pb sites, although that of P1 of Bi does not change at the pressure range between 7-20 GPa.

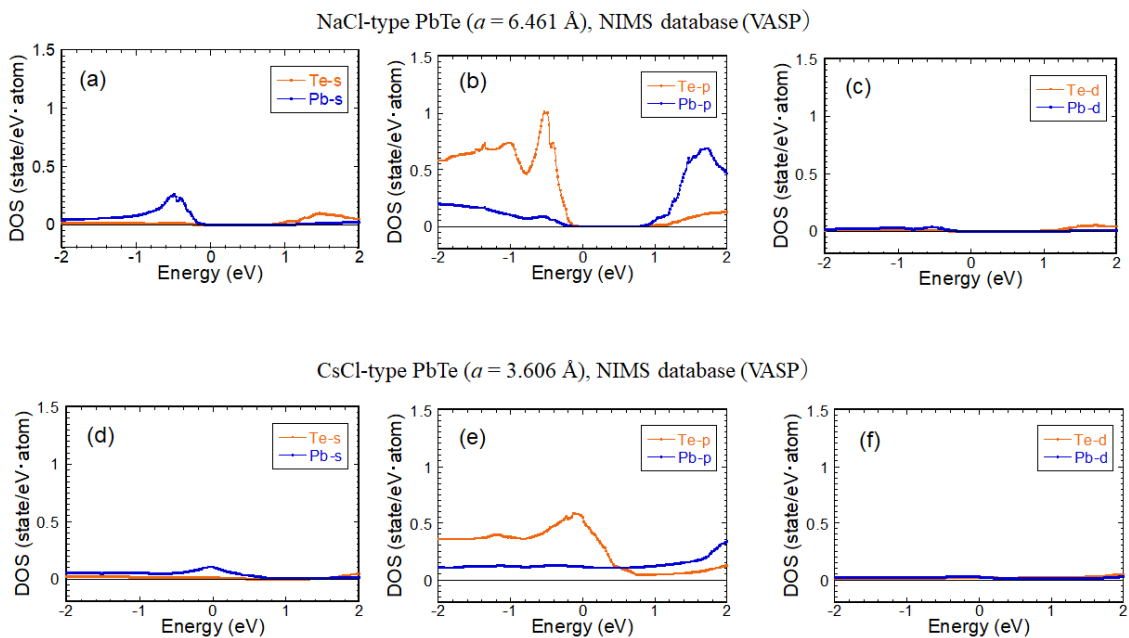

Fig. S8. (a–c) Calculated s, p, d density of states (DOS) of PbTe with a NaCl-type structure. (d–f) Calculated s, p, d density of states (DOS) of PbTe with a CsCl-type structure. The electronic density of states was calculated by CompES-X, NIMS database (https://compes-x.nims.go.jp/index.html) on 29 September 2021.